\documentclass[journal=jacsat,manuscript=article]{achemso}

\usepackage{siunitx}

\author{Michael Foltýn}
\affiliation[CEITEC]
{Brno University of Technology, Central European Institute of Technology, Purkyňova 123, Brno,
612 00, Czech Republic}

\author{Tomáš Šikola}
\affiliation[CEITEC]
{Brno University of Technology, Central European Institute of Technology, Purkyňova 123, Brno,
612 00, Czech Republic}
\alsoaffiliation[UFI]
{Brno University of Technology, Faculty of Mechanical Engineering, Institute of Physical Engineering, Technická 2, Brno, 616 69, Czech Republic}

\author{Michal Horák}
\affiliation[CEITEC]
{Brno University of Technology, Central European Institute of Technology, Purkyňova 123, Brno,
612 00, Czech Republic}
\email{michal.horak2@ceitec.vutbr.cz}

\title[Bismuth plasmonic antennas] {Bismuth plasmonic antennas}

\begin{document}

\begin{abstract}
  
Bismuth is a particularly promising alternative plasmonic metal because of its theoretically predicted wide spectral bandwidth. In this study, we experimentally demonstrated the correlation between the shape and size of individual bismuth plasmonic antennas and their optical properties. To this end, we employed a combination of scanning transmission electron microscopy and electron energy loss spectroscopy. Bar-shaped and bowtie bismuth plasmonic antennas of various sizes were fabricated by focused ion beam lithography of a polycrystalline bismuth thin film. Our experimental findings demonstrate that these antennas support localised surface plasmon resonances and their dipole modes can be tuned through their size from the near-infrared to the entire visible spectral region. Furthermore, our findings demonstrate that bismuth exhibits a plasmon dispersion relation that is nearly identical to that of gold while maintaining its plasmonic performance even at higher plasmon energies, thus rendering it a promising low-cost alternative to gold.
  
\end{abstract}

\section{Introduction}

Collective oscillations of free electrons in metallic nanoparticles, called localised surface plasmon resonances (LSPR), are known to enhance the local electromagnetic field in the vicinity of nanoparticles \cite{10.1038/nmat2630}. These extraordinary properties of plasmonic nanostructures, often called plasmonic antennas, have been used in various applications, including biosensing \cite{10.1021/acssensors.8b00315, 10.1515/nanoph-2023-0317}, catalysis \cite{10.1021/jacs.3c14586}, and ultrathin optical elements \cite{10.1038/lsa.2013.28, 10.1364/oe.506069}. Gold has been a material of choice and subject of study for many years in the context of plasmonic applications. However, the strong damping of LSPRs at energies above \SI{2}{\electronvolt}, caused by the gold interband transitions, limits the use of gold plasmonic antennas to the near infrared and a part of the visible spectral region \cite{10.1021/nn102166t}. This limitation has prompted the exploration of alternative non-noble plasmonic metals, such as gallium \cite{10.1021/acs.jpclett.3c00094}, magnesium \cite{10.1021/acs.jpcc.2c01944}, and potassium \cite{10.1021/acs.nanolett.3c02054}. Another material that has been theoretically predicted to offer a spectral interval wider than the visible region is bismuth \cite{10.1039/C3CP43856B}.

The low effective mass of free electrons and the dielectric function of bismuth suggest that it is an attractive plasmonic material suitable for plasmonics spanning from the near-infrared to the ultraviolet spectral region \cite{10.1103/PhysRevLett.98.076603, 10.1155/2018/3250932}. Furthermore, the extraordinary properties of bismuth, including quantum confinement \cite{10.1063/1.2192624}, temperature-induced metal-to-semiconductor transition \cite{10.1088/0957-4484/21/40/405701}, and high values of the Seebeck coefficient \cite{10.1002/anie.201005023}, when combined with its plasmonic performance, have the potential to yield new applications. Despite theoretical predictions of plasmonic activity in bismuth\cite{10.1039/C3CP43856B}, experimental research in this field has been limited to investigating circular or spherical bismuth nanostructures using far-field optical spectroscopy \cite{10.1021/acs.inorgchem.1c02621, 10.1016/j.photonics.2022.101058, 10.1021/jp3065882, 10.1364/ol.45.000686, 10.1002/adom.202302130}. The primary challenge of the method is that it measures the overall response of an ensemble of nanoparticles rather than individual ones. This is because real samples always contain nanoparticles of various sizes, resulting in plasmonic resonances that overlap in the measured spectrum. Consequently, it becomes impossible to isolate the contribution of a single nanoparticle \cite{10.1155/2013/313081, 10.1021/ac502053s}. As a result, the exploration of the spectral tunability of LSPRs in individual bismut nanostructures as a function of their size using previously employed techniques is rendered unfeasible.

In this work, we present a study that uses electron energy-loss spectroscopy in a scanning transmission electron microscope (STEM-EELS) \cite{10.1103/revmodphys.82.209} to address the optical response of individual bismuth plasmonic antennas. We show the spectral tunability of the dipole LSPR modes over the near-infrared and visible spectral range and correlate it with the size of bismuth nanostructures.

\section{Results and discussion}

\begin{figure}
    \centering
    \includegraphics[width=1\linewidth]{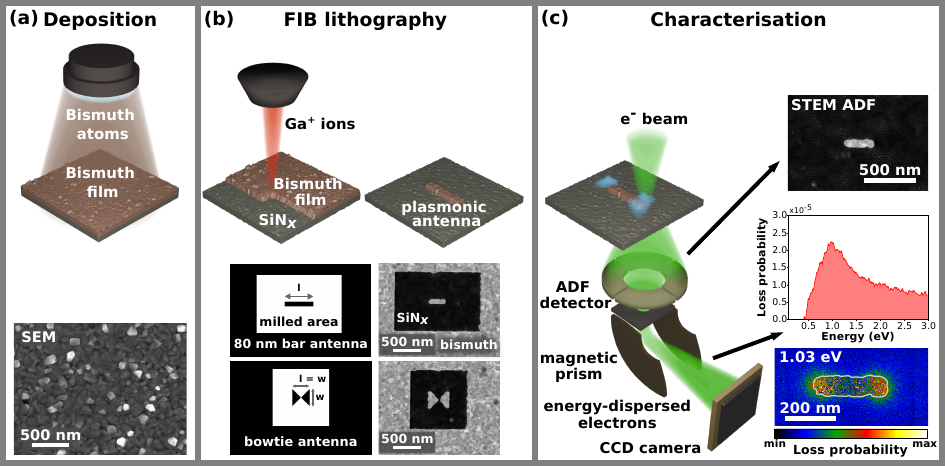}
    \caption{Schematic workflow for the fabrication and characterisation of bismuth plasmonic antennas: (a) bismuth thin films were deposited by magnetron sputtering on a silicon nitride membrane, (b) bar-shaped and bowtie bismuth plasmonic antennas were fabricated by FIB lithography, and (c) their morphology was captured by STEM ADF micrographs and LSPRs were measured by STEM EELS.}
    \label{Fig1}
\end{figure}

Bismuth plasmonic antennas were prepared using a standard focused ion beam (FIB) lithography process \cite{10.1038/s41598-018-28037-1} and characterised using STEM-EELS \cite{10.1016/j.ultramic.2020.113044}. The schematic workflow for the fabrication and characterisation of bismuth plasmonic antennas is shown in Figure~\ref{Fig1}. \SI{30}{\nano\meter} thick bismuth films were deposited on commercially available silicon nitride membranes by magnetron sputtering (Figure~\ref{Fig1}a). The micrograph obtained by scanning electron microscopy (SEM) shows the deposited bismuth layer with polygonal grains. The cross-sectional view of a lamella cut off from the sample shows a pronounced roughness of the bismuth polycrystalline layer with no signs of oxidation (see Figure~S1 in Supplementary Information). To further assess the oxidation resistance of the bismuth thin films when exposed to air, we employed a series of diffraction experiments while the diffractograms measured approximately half a year after deposition exhibited no indications of any bismuth oxide crystal phases (see Figure~S2 in Supplementary Information). Plasmonic antennas were fabricated from bismuth thin films by FIB lithography (Figure~\ref{Fig1}b). We targeted bar-shaped and bowtie bismuth plasmonic antennas. The width of the bar-shaped antennas is \SI{80}{\nano\meter} and their length varies from 100 to \SI{500}{\nano\meter}. Bowtie antennas have a wing angle of \SI{90}{\degree} and their total length, which is equal to their width, ranges from 130 to \SI{620}{\nano\meter}. FIB lithography patterns with marked length of the bar-shaped antenna and width of the bowtie antenna are shown in Figure~\ref{Fig1}b as well as STEM annular dark field (ADF) micrographs of \SI{286}{\nano\meter} long bar and \SI{288}{\nano\meter} wide bowtie antenna. Figure~\ref{Fig1}c shows a set-up of electron energy loss spectroscopy used to measure the EELS of individual bismuth nanostructures with analysis of the \SI{294}{\nano\meter} long bar antenna including STEM ADF micrograph, background subtracted electron energy loss spectrum integrated over the left corner of the antenna with a peak at \SI{1.03}{\electronvolt} corresponding to a longitudinal dipole LSPR mode, and the loss probability map at peak energy (\SI{1.03}{\electronvolt}) showing the spatial distribution of this mode with two maxima at the corners of the antenna.

\begin{figure}
    \centering
    \includegraphics[width=1\linewidth]{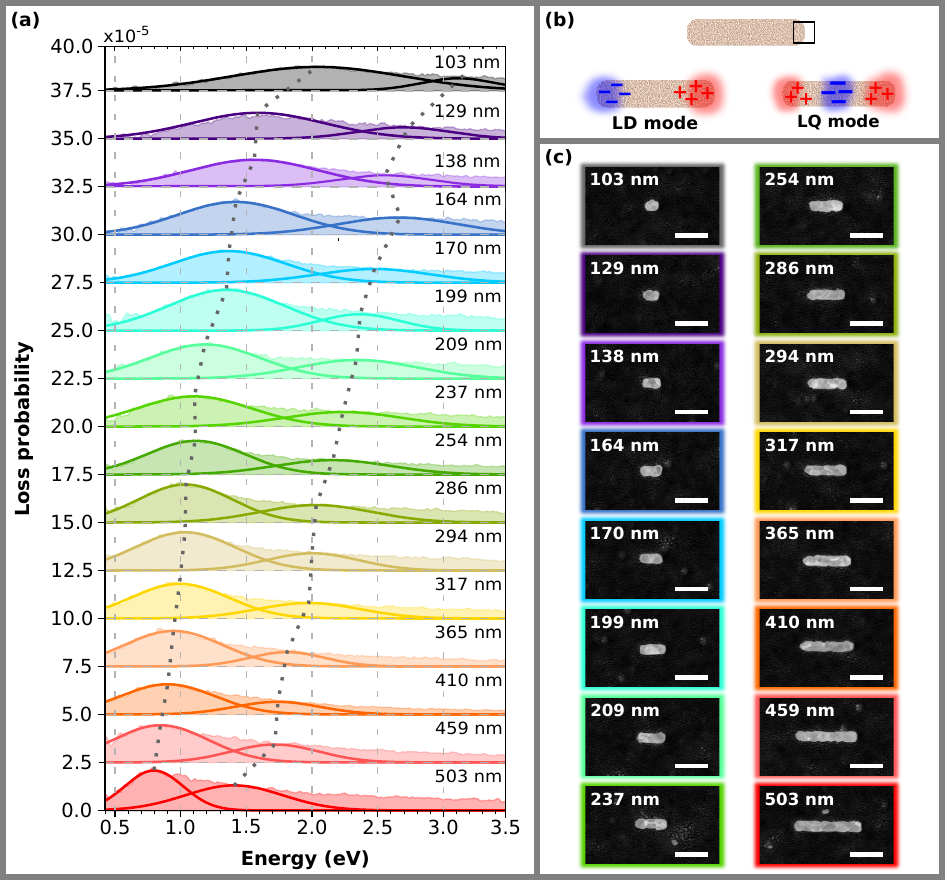}
    \caption{EELS analysis of bar antennas: (a) Measured EEL spectra (further fitted with two Gaussians) of bar antennas with the length ranging from \qtyrange{103}{503}{\nano\meter}. The first peak in every spectrum corresponds to the longitudinal dipole (LD) mode and the second to the longitudinal quadrupole (LQ) mode. Dashed lines are guides for the eye and follow the energy of LD and LQ modes that increases with the decreasing length of the bar antenna. (b) A schema of the LD and LQ mode with marked area at the edge of the antenna where the EEL spectra were collected. (c) STEM ADF micrographs of individual analysed bar antennas. The length of the scalebars is \SI{250}{\nano\meter}.}
    \label{Fig2}
\end{figure}

First, we inspect the set of 16 bar-shaped antennas with a length from \SI{103}{\nano\meter} to \SI{503}{\nano\meter}. The results are summarised in Figure~\ref{Fig2}. Figure~\ref{Fig2}a presents EEL spectra measured on the edges of fabricated bar antennas (the integration area is marked in Figure~\ref{Fig2}b by the black rectangle), whose STEM ADF micrographs are depicted in Figure~\ref{Fig2}c. These EEL spectra exhibit pronounced peaks in the energy region \qtyrange{0.5}{3.5}{\electronvolt} corresponding to the longitudinal dipole (LD) and longitudinal quadrupole (LQ) LSPR mode, which are schematically shown in Figure~\ref{Fig2}b. The identification of modes was performed with the help of numerical simulations and loss probability maps. A comparison of the EELS experiment and the theory for the \SI{400}{\nano\meter} bar is shown in Figure~S3 and the loss probability maps of all bar antennas are shown in Figure~S4 in Supplementary Information. The measured EEL spectra were fitted by two Gaussian curves to extract the characteristic parameters of the individual observed plasmon peaks. It includes the peak position corresponding to the energy of the respective LSPR mode, the loss probability maximum, and the full width at half-maximum (FWHM) of the peak. The loss probability maxima obtained for both LD and LQ modes remain approximately the same for all antenna lengths, with an observable decrease for the shortest antennas.

In the following, we will focus on the LD mode. The highest loss probability of $2.29 \times 10^{-5}$ and $2.25 \times 10^{-5}$ was observed in \SI{503}{\nano\meter} and \SI{199}{\nano\meter} long antennas, respectively, while the lowest loss probability of $1.34 \times 10^{-5}$ was observed in the \SI{103}{\nano\meter} long bar antenna. Similarly, the FWHM of fitted plasmon peaks remains constant, with an observable increase for the shortest antennas. The lowest FWHM of \SI{0.22}{\electronvolt} was observed in the \SI{503}{\nano\meter} long antenna. For all remaining antennas longer than \SI{230}{\nano\meter}, the FWHM remains below \SI{0.37}{\electronvolt}. For shorter antennas, the FWHM increases to \SI{0.65}{\electronvolt} for the shortest antenna. The origin of the observed broadening and the decrease in the loss probability for the shortest antennas can be attributed to the dielectric properties of bismuth. For bar antennas longer than \SI{200}{\nano\meter}, the plasmonic response such as the peak intensity and the FWHM remains comparable, offering stable plasmonic performance regardless of the antenna dimension. Consequently, bismuth bar antennas represent a vivid plasmonic system tunable from the visible to the near-infrared spectral region. 

\begin{figure}
    \centering
    \includegraphics[width=1\linewidth]{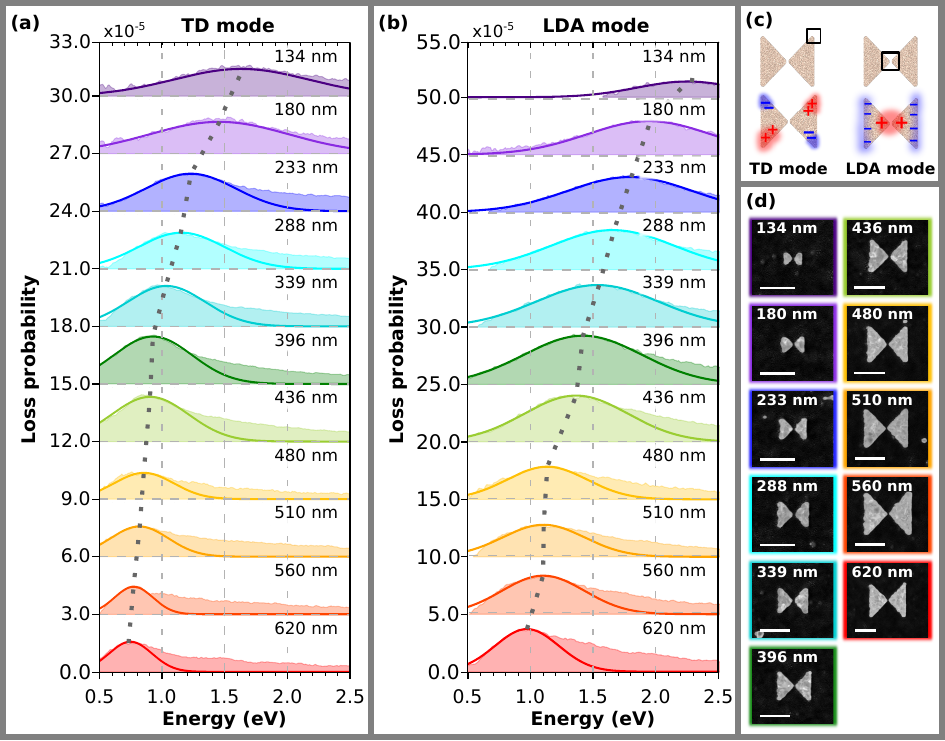}
    \caption{EELS analysis of bowtie antennas: (a,b) Measured EEL spectra (further fitted with a Gaussian) from the outer corners (a), where the peak corresponds to the transverse dipole (TD) mode, and gaps (b), where the peak corresponds to the longitudinal dipole antibonding (LDA) mode, of bowtie antennas with the width ranging from \qtyrange{134}{620}{\nano\meter}. Dashed lines are guides for the eye and follow the energy of TD and LDA modes that increases with the decreasing width of the bowtie antenna. (c) Schematic depiction of the TD and LDA mode with marked areas where the EEL spectra were collected. (d) STEM ADF micrographs of the analysed bowtie antennas. The length of the scalebars is \SI{400}{\nano\meter}.}
    \label{Fig3}
\end{figure}

Second, we inspect the set of 11 bowtie antennas with a width from \SI{134}{\nano\meter} to \SI{620}{\nano\meter}. The results are summarised in Figure~\ref{Fig3}. To characterise the plasmonic performance, we studied the EEL spectra at the bowtie corners (Figure~\ref{Fig3}a) and at the gap between its two wings (Figure~\ref{Fig3}b). The integration area where the spectra were collected is depicted in Figure~\ref{Fig3}c together with a schema of the transverse dipole (TD) mode with the maxima at the outer corners of the bowtie and the longitudinal dipole antibonding (LDA) mode with the maximum at the gap of the bowtie. The peaks in the measured EEL spectra and their respective plasmon modes were identified with the help of numerical simulations and loss probability maps. A comparison of the EELS experiment and the theory for the \SI{400}{\nano\meter} bowtie is shown in Figure~S5 and the loss probability maps TD and LDA modes of all bowtie antennas are shown in Figure~S6 in Supplementary Information. The peak energy, loss probability maxima, and FWHM of the TD and LDA modes were obtained by fitting the measured spectra with Gaussian curves. In the case of the TD mode, the loss probability is the highest for antennas with widths from \qtyrange{180}{480}{\nano\meter}. The highest value of $2.60 \times 10^{-5}$ is reached by the \SI{396}{\nano\meter} bowtie. For bowties below and above the specified antenna width range, the loss probability decreases. The lowest observed loss probability of $1.12 \times 10^{-5}$ was measured in the \SI{480}{\nano\meter} antenna. The FWHM of the TD plasmon peaks increases with decreasing antenna width. The lowest FWHM of \SI{0.17}{\electronvolt} was evaluated in the \SI{620}{\nano\meter} bowtie, whereas the highest FWHM of \SI{0.56}{\electronvolt} was observed in the \SI{180}{\nano\meter} antenna. Antennas with widths between 233 and \SI{480}{\nano\meter} exhibit TD plasmon peaks with comparable FWHM values between 0.30 and \SI{0.36}{\electronvolt}. In the case of the LDA mode, the loss probability increases for larger antennas and culminates at $4.40 \times 10^{-5}$ for the \SI{396}{\nano\meter} bowtie. The lowest loss probability of $1.43 \times 10^{-5}$ was measured in the smallest (\SI{134}{\nano\meter}) antenna. The FWHM increases with decreasing antenna width, from the lowest value of \SI{0.25}{\electronvolt} in the largest antenna to \SI{0.50}{\electronvolt} measured in the second smallest bowtie.

\begin{figure}
    \centering
    \includegraphics[width=1\linewidth]{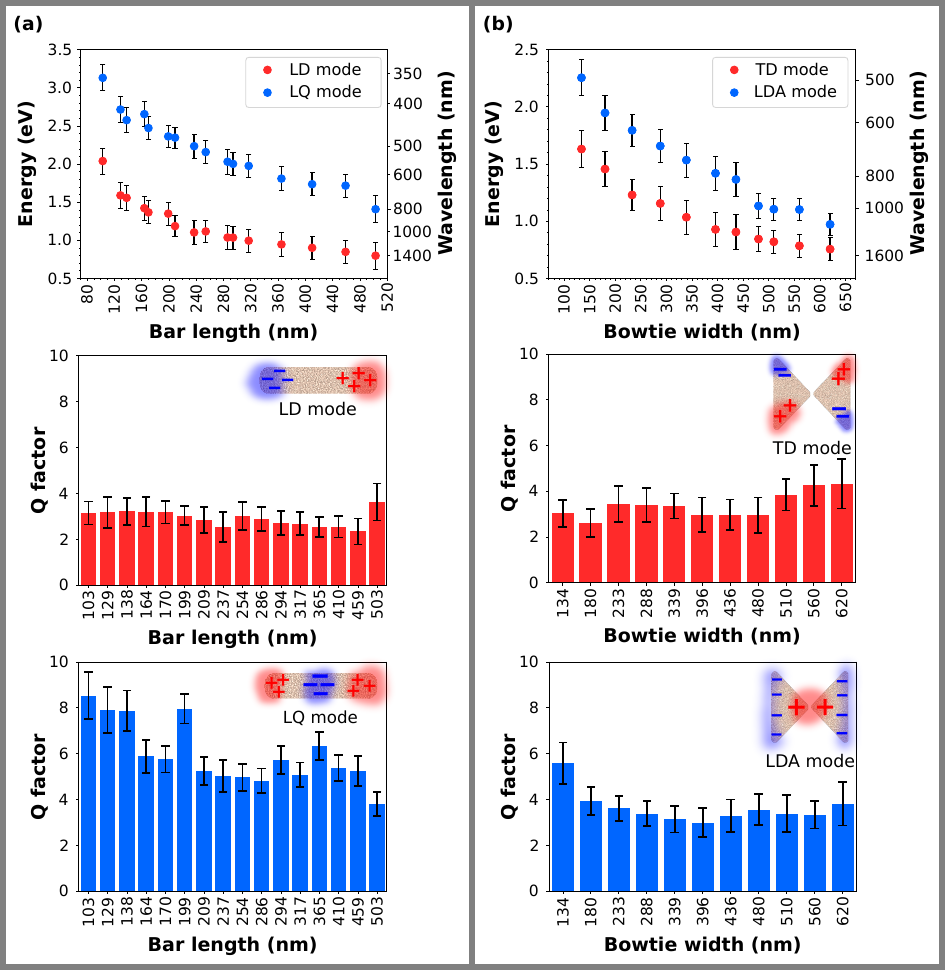}
    \caption{Spectral tunability of bismuth plasmonic antennas: (a) Plasmon energy and Q factors of the LD and LQ modes extracted from measured EEL spectra shown in Figure~\ref{Fig2} as a function of the length of the bar antennas. (b) Plasmon energy and Q factors of the TD and LDA modes extracted from measured EEL spectra shown in Figure~\ref{Fig3} as a function of the width of the bowtie antennas.}
    \label{Fig4}
\end{figure}

Third, we evaluate the spectral tunability of bismuth plasmonic antennas. The plasmon resonances in both the bowtie and bar antennas fall within the near-infrared to visible part of the spectrum.

In the case of bar antennas (Figure~\ref{Fig4}a), the LD mode energy covers the interval from \SI{0.78}{\electronvolt} (corresponding to \SI{1550}{\nano\meter} in wavelength) for the longest \SI{503}{\nano\meter} bar to \SI{2.03}{\electronvolt} (\SI{610}{\nano\meter} in wavelength) for the shortest \SI{103}{\nano\meter} antenna. The LQ mode energy then covers the energy interval from \SI{1.41}{\electronvolt} (\SI{879}{\nano\meter} in wavelength) for the longest to \SI{3.13}{\electronvolt} (\SI{396}{\nano\meter} in wavelength) for the shortest antenna. The Q factor of the LD mode remains constant for all antenna lengths and fluctuates around the value of 3. In the case of the LQ mode, the Q factors are higher than those for the LD modes. For the shortest antenna, the Q factor of the LQ mode is the highest, reaching a value of 8.5. With increasing antenna width, the Q factor values decrease, reaching a minimum of 3.8 for the longest bar. 

In the case of bowties (Figure~\ref{Fig4}b), the energy of the TD mode ranges from \SI{0.76}{\electronvolt} (\SI{1631}{\nano\meter} in wavelength) for the largest \SI{620}{\nano\meter} bowtie to \SI{1.63}{\electronvolt} (\SI{760}{\nano\meter} in wavelength) for the smallest \SI{134}{\nano\meter} bowtie. The LDA mode energy interval covers the interval from \SI{0.97}{\electronvolt} (\SI{1278}{\nano\meter} in wavelength) for the largest bowtie to \SI{2.25}{\electronvolt} (\SI{551}{\nano\meter} in wavelength) for the smallest bowtie. The Q factors of the TD modes fluctuate around the value of 3 for all bowtie antenna widths, with the exception of antennas with widths greater than \SI{500}{\nano\meter}, for which the Q factor increases above 4. In the case of LDA modes, the Q factors are higher, and they fluctuate constantly around the value of 3.5 except for the smallest antennas (widths around 200 nm and smaller), where the Q factors increase, reaching the maximum of 5.6 for the smallest bowtie antenna. 

In summary, the dependence of the plasmon energy on the antenna width proves the suitability of bismuth to cover the near-infrared spectral region and, with sufficiently small (below \SI{100}{\nano\meter}) nanostructures, even the entire visible region while maintaining its performance across the spectral bandwidth. However, the use of lithographically fabricated nanostructures of the two tested geometries is not capable of supporting LSPR in the ultraviolet spectral region, as very small structures (below \SI{50}{\nano\meter} in size) would be necessary. Such small structures can be achieved, for example, by chemical synthesis.

\begin{figure}
    \centering
    \includegraphics[width=0.6\linewidth]{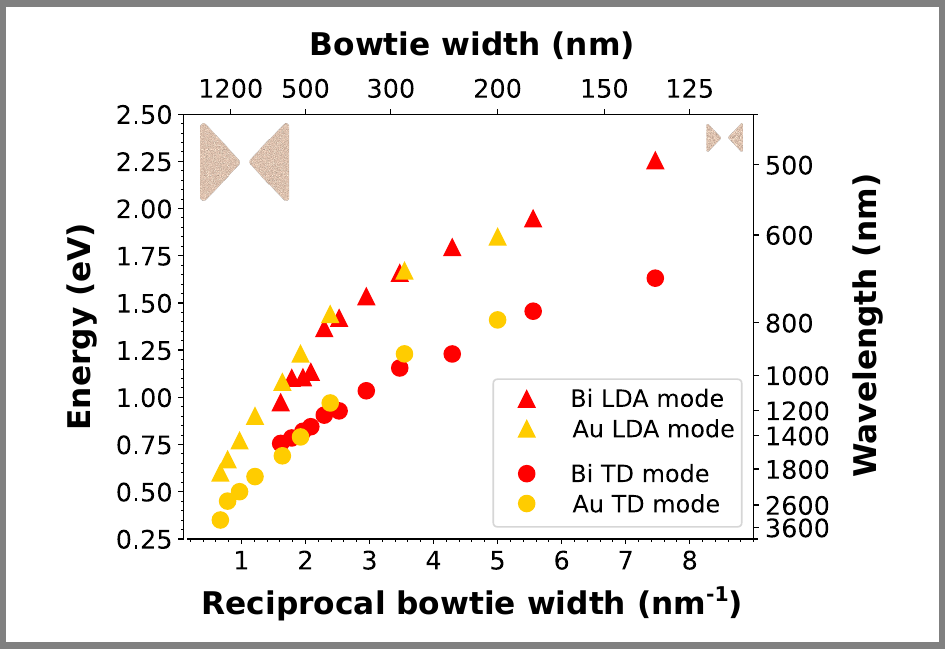}
    \caption{Dispersion relation of LSPR in bismuth and gold bowties: Energy of the TD and LDA modes in bismuth (data from Figure~\ref{Fig4}b) and gold (data from Ref. \cite{10.1515/nanoph-2019-0326}) bowties is plotted as a function of the reciprocal antenna width. The overlap of the dependencies for both materials suggests the full substitutability of gold by bismuth in plasmonic applications.}
    \label{Fig5}
\end{figure}

Finally, we compare the plasmonic properties of bismuth with the plasmonic properties of gold. A direct comparison of the plasmon energy of the TD and LDA modes in bismuth and gold bowties is shown in Figure~\ref{Fig5}. The dispersion relations for these two modes in bismuth bowties overlap with the corresponding dispersion relations for gold bowties. The observed overlap of the dispersion relations suggests that bismuth can be considered an alternative material to gold. Despite the fact that the Q factors of the bismuth antennas are marginally lower than those of gold (see Figure~S7 in Supplementary Information), this disadvantage is counterbalanced by their consistent performance even at higher plasmon energies. Furthermore, the significantly lower cost of bismuth antennas enhances their applicability. 

It is imperative to consider two other pivotal factors that influence the suitability of various plasmonic materials: their biocompatibility and chemical stability under ambient conditions, such as their resistance to oxidation when exposed to air. In both cases, bismuth exhibited properties comparable to those of gold. The present study employed a series of diffraction experiments using both X-ray and electron diffraction to assess the oxidation resistance of bismuth thin films when exposed to air. The diffractograms measured approximately half a year after deposition exhibited no indications of any bismuth oxide crystal phases, despite prolonged exposure of the bismuth film to ambient conditions. Consequently, bismuth in the form of thin polycrystalline layers appears to be chemically stable and resistant to oxidation. The diffractograms are included as Figure~S2 in Supplementary Information.

In terms of its toxicity, bismuth is generally considered biocompatible. The only bodily harm caused by bismuth is inflammation of the lungs after inhalation of fine bismuth powder, which is caused by mechanical irritation of the tissue. However, when bismuth is introduced into the body in other forms, no negative effects on mammals have been observed, even at doses as high as \SI{1000}{\milli\gram} per \SI{1}{\kilo\gram} of body mass \cite{10.1539/joh.47.242, 10.1039/D0CS00031K}. Consequently, bismuth emerges as a promising alternative to gold, offering cost-effectiveness, chemical stability, biocompatibility, and enhanced plasmon energy tunability, albeit at the expense of reduced plasmon resonance intensity. 

\section{Conclusion}

In conclusion, we have fabricated bar-shaped and bowtie bismuth plasmonic antennas using a standard focused ion beam lithography of a polycrystalline bismuth thin film and characterised them using STEM-EELS. This approach enables the study of plasmonic properties at the single-particle level and the exploration of the spectral tunability of localised surface plasmon resonances in individual bismuth nanostructures as a function of their size. The spectral tunability of single modes over the near-infrared and visible spectral range has been demonstrated, and a correlation with the size of bismuth nanostructures has been established.

Our experimental results therefore demonstrate that bismuth is a suitable and cost-effective material for plasmonic applications. The dipole modes in the explored nanostructures are tunable from the near-infrared spectral region to the entire visible region. Furthermore, the plasmon resonances exhibited by these structures are found to be stable over the entire plasmon energy interval. Moreover, we have shown that the dependence of the plasmon energy on the antenna size for gold and bismuth is highly congruent, thereby establishing bismuth as a viable alternative to gold. This is further underscored by the observation that bismuth also covers the energies above \SI{2}{\electronvolt}. Additionally, the lower cost of bismuth, together with its biocompatibility and resistance to oxidation, make it a suitable candidate for use, especially in industrial and large-scale plasmonic applications.

\section{Methods}
\subsection{Metal deposition}
A \SI{30}{\nano\meter} thick bismuth thin film was deposited on a \SI{30}{\nano\meter} silicon nitride membrane (by Agar Scientific) with lateral dimensions of $250 \times 250$\SI{}{\micro\meter\squared} by DC magnetron sputtering. We used Magnetron Sputtering System BESTEC with the following parameters: chamber pressure $8 \cdot 10^{-4}$\,mbar, sample rotation 5\,rpm, argon gas flux 15\,sccm, argon ion energy \SI{310}{\electronvolt}, and total current of argon ions \SI{25}{\milli\ampere} resulting in the deposition rate of \SI{0.45}{\angstrom\per\second}. 

\subsection{FIB lithography}
FIB lithography of the polycrystalline bismuth thin film was performed using FEI Helios by gallium ions with an energy of \SI{30}{\kilo\electronvolt} and an ion beam current of \SI{2}{\pico\ampere}. We note that the highest available beam energy and the lowest available beam current are optimised for the best spatial resolution of the milling.

\subsection{EELS measurement}
EELS measurements were carried out in a TEM FEI Titan equipped with a GIF Quantum spectrometer operated at \SI{120}{\kilo\electronvolt} in the scanning monochromated mode with the convergence semi-angle set to 10\,mrad and the collection semi-angle set to 11.4\,mrad. The probe current was adjusted to around \SI{100}{\pico\ampere}. The dispersion of the spectrometer was set to \SI{0.01}{\electronvolt} per channel and the FWHM of the zero-loss peak was around \SI{0.15}{\electronvolt}. The acquisition time was adjusted to use the maximal intensity range of the CCD camera in the spectrometer and avoid its overexposure. EEL spectra were integrated over rectangular areas at the edges of the nanostructures where the LSPR is significant. They were further divided by the integral intensity of the zero-loss peak to transform the measured counts into a quantity proportional to the loss probability, background subtracted by subtracting the EEL spectrum of a pure silicon nitride membrane, and fitted by Gaussians.

\subsection{Numerical simulations}
Numerical simulations of EELS spectra were performed using the MNPBEM toolbox \cite{10.1016/j.cpc.2015.03.023} based on the boundary element method. The dielectric function of bismuth was taken from Ref. \cite{10.1063/1.3243762} and the surrounding dielectric constant was set to 1.6 to approximate the effect of the silicon nitride membrane substrate. The \SI{120}{\kilo\electronvolt} electron beam was positioned \SI{5}{\nano\meter} outside the antenna. The loss probability density was further recalculated to the loss probability at energy intervals of \SI{0.01}{\electronvolt} corresponding to the dispersion of the spectrometer in the experiment.

\begin{acknowledgement}

This work is supported by the project QM4ST (project No. CZ.02.01.01/00/22\_008/0004572) by OP JAK, call Excellent Research, project Czech-NanoLab by MEYS CR (project No. LM2023051), and Brno University of Technology (project No. FSI-S-23-8336). M.F. acknowledges the support of the Brno Ph.D. talent scholarship.

\end{acknowledgement}


\bibliography{reference}

\providecommand{\latin}[1]{#1}
\makeatletter
\providecommand{\doi}
  {\begingroup\let\do\@makeother\dospecials
  \catcode`\{=1 \catcode`\}=2 \doi@aux}
\providecommand{\doi@aux}[1]{\endgroup\texttt{#1}}
\makeatother
\providecommand*\mcitethebibliography{\thebibliography}
\csname @ifundefined\endcsname{endmcitethebibliography}  {\let\endmcitethebibliography\endthebibliography}{}
\begin{mcitethebibliography}{32}
\providecommand*\natexlab[1]{#1}
\providecommand*\mciteSetBstSublistMode[1]{}
\providecommand*\mciteSetBstMaxWidthForm[2]{}
\providecommand*\mciteBstWouldAddEndPuncttrue
  {\def\EndOfBibitem{\unskip.}}
\providecommand*\mciteBstWouldAddEndPunctfalse
  {\let\EndOfBibitem\relax}
\providecommand*\mciteSetBstMidEndSepPunct[3]{}
\providecommand*\mciteSetBstSublistLabelBeginEnd[3]{}
\providecommand*\EndOfBibitem{}
\mciteSetBstSublistMode{f}
\mciteSetBstMaxWidthForm{subitem}{(\alph{mcitesubitemcount})}
\mciteSetBstSublistLabelBeginEnd
  {\mcitemaxwidthsubitemform\space}
  {\relax}
  {\relax}

\bibitem[Schuller \latin{et~al.}(2010)Schuller, Barnard, Cai, Jun, White, and Brongersma]{10.1038/nmat2630}
Schuller,~J.~A.; Barnard,~E.~S.; Cai,~W.; Jun,~Y.~C.; White,~J.~S.; Brongersma,~M.~L. Plasmonics for extreme light concentration and manipulation. \emph{Nature Materials} \textbf{2010}, \emph{9}, 193–204\relax
\mciteBstWouldAddEndPuncttrue
\mciteSetBstMidEndSepPunct{\mcitedefaultmidpunct}
{\mcitedefaultendpunct}{\mcitedefaultseppunct}\relax
\EndOfBibitem
\bibitem[Klinghammer \latin{et~al.}(2018)Klinghammer, Uhlig, Patrovsky, B\"{o}hm, Sch\"{u}tt, P\"{u}tz, Baraban, Eng, and Cuniberti]{10.1021/acssensors.8b00315}
Klinghammer,~S.; Uhlig,~T.; Patrovsky,~F.; B\"{o}hm,~M.; Sch\"{u}tt,~J.; P\"{u}tz,~N.; Baraban,~L.; Eng,~L.~M.; Cuniberti,~G. Plasmonic Biosensor Based on Vertical Arrays of Gold Nanoantennas. \emph{ACS Sensors} \textbf{2018}, \emph{3}, 1392–1400\relax
\mciteBstWouldAddEndPuncttrue
\mciteSetBstMidEndSepPunct{\mcitedefaultmidpunct}
{\mcitedefaultendpunct}{\mcitedefaultseppunct}\relax
\EndOfBibitem
\bibitem[Riley \latin{et~al.}(2023)Riley, Horák, Křápek, Healy, and Pacheco-Peña]{10.1515/nanoph-2023-0317}
Riley,~J.~A.; Horák,~M.; Křápek,~V.; Healy,~N.; Pacheco-Peña,~V. Plasmonic sensing using Babinet’s principle. \emph{Nanophotonics} \textbf{2023}, \emph{12}, 3895–3909\relax
\mciteBstWouldAddEndPuncttrue
\mciteSetBstMidEndSepPunct{\mcitedefaultmidpunct}
{\mcitedefaultendpunct}{\mcitedefaultseppunct}\relax
\EndOfBibitem
\bibitem[Yang \latin{et~al.}(2024)Yang, Jia, Hu, Zhao, Li, Ni, and Zhang]{10.1021/jacs.3c14586}
Yang,~Y.; Jia,~H.; Hu,~N.; Zhao,~M.; Li,~J.; Ni,~W.; Zhang,~C.-y. Construction of Gold/Rhodium Freestanding Superstructures as Antenna-Reactor Photocatalysts for Plasmon-Driven Nitrogen Fixation. \emph{Journal of the American Chemical Society} \textbf{2024}, \emph{146}, 7734–7742\relax
\mciteBstWouldAddEndPuncttrue
\mciteSetBstMidEndSepPunct{\mcitedefaultmidpunct}
{\mcitedefaultendpunct}{\mcitedefaultseppunct}\relax
\EndOfBibitem
\bibitem[Ni \latin{et~al.}(2013)Ni, Ishii, Kildishev, and Shalaev]{10.1038/lsa.2013.28}
Ni,~X.; Ishii,~S.; Kildishev,~A.~V.; Shalaev,~V.~M. Ultra-thin, planar, Babinet-inverted plasmonic metalenses. \emph{Light: Science \& Applications} \textbf{2013}, \emph{2}, e72\relax
\mciteBstWouldAddEndPuncttrue
\mciteSetBstMidEndSepPunct{\mcitedefaultmidpunct}
{\mcitedefaultendpunct}{\mcitedefaultseppunct}\relax
\EndOfBibitem
\bibitem[Rovenská \latin{et~al.}(2023)Rovenská, Ligmajer, Idesová, Kepič, Liška, Chochol, and Šikola]{10.1364/oe.506069}
Rovenská,~K.; Ligmajer,~F.; Idesová,~B.; Kepič,~P.; Liška,~J.; Chochol,~J.; Šikola,~T. Structural color filters with compensated angle-dependent shifts. \emph{Optics Express} \textbf{2023}, \emph{31}, 43048\relax
\mciteBstWouldAddEndPuncttrue
\mciteSetBstMidEndSepPunct{\mcitedefaultmidpunct}
{\mcitedefaultendpunct}{\mcitedefaultseppunct}\relax
\EndOfBibitem
\bibitem[Zorić \latin{et~al.}(2011)Zorić, Z\"{a}ch, Kasemo, and Langhammer]{10.1021/nn102166t}
Zorić,~I.; Z\"{a}ch,~M.; Kasemo,~B.; Langhammer,~C. Gold, Platinum, and Aluminum Nanodisk Plasmons: Material Independence, Subradiance, and Damping Mechanisms. \emph{ACS Nano} \textbf{2011}, \emph{5}, 2535–2546\relax
\mciteBstWouldAddEndPuncttrue
\mciteSetBstMidEndSepPunct{\mcitedefaultmidpunct}
{\mcitedefaultendpunct}{\mcitedefaultseppunct}\relax
\EndOfBibitem
\bibitem[Horák \latin{et~al.}(2023)Horák, Čalkovský, Mach, Křápek, and Šikola]{10.1021/acs.jpclett.3c00094}
Horák,~M.; Čalkovský,~V.; Mach,~J.; Křápek,~V.; Šikola,~T. Plasmonic Properties of Individual Gallium Nanoparticles. \emph{The Journal of Physical Chemistry Letters} \textbf{2023}, \emph{14}, 2012–2019\relax
\mciteBstWouldAddEndPuncttrue
\mciteSetBstMidEndSepPunct{\mcitedefaultmidpunct}
{\mcitedefaultendpunct}{\mcitedefaultseppunct}\relax
\EndOfBibitem
\bibitem[Hopper \latin{et~al.}(2022)Hopper, Boukouvala, Asselin, Biggins, and Ringe]{10.1021/acs.jpcc.2c01944}
Hopper,~E.~R.; Boukouvala,~C.; Asselin,~J.; Biggins,~J.~S.; Ringe,~E. Opportunities and Challenges for Alternative Nanoplasmonic Metals: Magnesium and Beyond. \emph{The Journal of Physical Chemistry C} \textbf{2022}, \emph{126}, 10630–10643\relax
\mciteBstWouldAddEndPuncttrue
\mciteSetBstMidEndSepPunct{\mcitedefaultmidpunct}
{\mcitedefaultendpunct}{\mcitedefaultseppunct}\relax
\EndOfBibitem
\bibitem[Gao \latin{et~al.}(2023)Gao, Wildenborg, Kocoj, Liu, Sheofsky, Rawashdeh, Qu, Guo, Suh, and Yang]{10.1021/acs.nanolett.3c02054}
Gao,~Z.; Wildenborg,~A.; Kocoj,~C.~A.; Liu,~E.; Sheofsky,~C.; Rawashdeh,~A.; Qu,~H.; Guo,~P.; Suh,~J.~Y.; Yang,~A. Low-Loss Plasmonics with Nanostructured Potassium and Sodium–Potassium Liquid Alloys. \emph{Nano Letters} \textbf{2023}, \emph{23}, 7150–7156\relax
\mciteBstWouldAddEndPuncttrue
\mciteSetBstMidEndSepPunct{\mcitedefaultmidpunct}
{\mcitedefaultendpunct}{\mcitedefaultseppunct}\relax
\EndOfBibitem
\bibitem[McMahon \latin{et~al.}(2013)McMahon, Schatz, and Gray]{10.1039/C3CP43856B}
McMahon,~J.~M.; Schatz,~G.~C.; Gray,~S.~K. Plasmonics in the ultraviolet with the poor metals Al, Ga, In, Sn, Tl, Pb, and Bi. \emph{Phys. Chem. Chem. Phys.} \textbf{2013}, \emph{15}, 5415–5423\relax
\mciteBstWouldAddEndPuncttrue
\mciteSetBstMidEndSepPunct{\mcitedefaultmidpunct}
{\mcitedefaultendpunct}{\mcitedefaultseppunct}\relax
\EndOfBibitem
\bibitem[Behnia \latin{et~al.}(2007)Behnia, Méasson, and Kopelevich]{10.1103/PhysRevLett.98.076603}
Behnia,~K.; Méasson,~M.-A.; Kopelevich,~Y. Nernst Effect in Semimetals: The Effective Mass and the Figure of Merit. \emph{Physical Review Letters} \textbf{2007}, \emph{98}\relax
\mciteBstWouldAddEndPuncttrue
\mciteSetBstMidEndSepPunct{\mcitedefaultmidpunct}
{\mcitedefaultendpunct}{\mcitedefaultseppunct}\relax
\EndOfBibitem
\bibitem[Tian and Toudert(2018)Tian, and Toudert]{10.1155/2018/3250932}
Tian,~Y.; Toudert,~J. Nanobismuth: Fabrication, Optical, and Plasmonic Properties—Emerging Applications. \emph{Journal of Nanotechnology} \textbf{2018}, \emph{2018}, 1–23\relax
\mciteBstWouldAddEndPuncttrue
\mciteSetBstMidEndSepPunct{\mcitedefaultmidpunct}
{\mcitedefaultendpunct}{\mcitedefaultseppunct}\relax
\EndOfBibitem
\bibitem[Wang \latin{et~al.}(2006)Wang, Kim, Kim, and Kim]{10.1063/1.2192624}
Wang,~Y.~W.; Kim,~J.~S.; Kim,~G.~H.; Kim,~K.~S. Quantum size effects in the volume plasmon excitation of bismuth nanoparticles investigated by electron energy loss spectroscopy. \emph{Applied Physics Letters} \textbf{2006}, \emph{88}\relax
\mciteBstWouldAddEndPuncttrue
\mciteSetBstMidEndSepPunct{\mcitedefaultmidpunct}
{\mcitedefaultendpunct}{\mcitedefaultseppunct}\relax
\EndOfBibitem
\bibitem[Lee \latin{et~al.}(2010)Lee, Ham, Jeon, Noh, and Lee]{10.1088/0957-4484/21/40/405701}
Lee,~S.; Ham,~J.; Jeon,~K.; Noh,~J.-S.; Lee,~W. Direct observation of the semimetal-to-semiconductor transition of individual single-crystal bismuth nanowires grown by on-film formation of nanowires. \emph{Nanotechnology} \textbf{2010}, \emph{21}, 405701\relax
\mciteBstWouldAddEndPuncttrue
\mciteSetBstMidEndSepPunct{\mcitedefaultmidpunct}
{\mcitedefaultendpunct}{\mcitedefaultseppunct}\relax
\EndOfBibitem
\bibitem[Son \latin{et~al.}(2010)Son, Park, Han, Kang, Park, Kim, Kim, Kim, and Hyeon]{10.1002/anie.201005023}
Son,~J.~S.; Park,~K.; Han,~M.; Kang,~C.; Park,~S.; Kim,~J.; Kim,~W.; Kim,~S.; Hyeon,~T. Large‐Scale Synthesis and Characterization of the Size‐Dependent Thermoelectric Properties of Uniformly Sized Bismuth Nanocrystals. \emph{Angewandte Chemie International Edition} \textbf{2010}, \emph{50}, 1363–1366\relax
\mciteBstWouldAddEndPuncttrue
\mciteSetBstMidEndSepPunct{\mcitedefaultmidpunct}
{\mcitedefaultendpunct}{\mcitedefaultseppunct}\relax
\EndOfBibitem
\bibitem[Leng \latin{et~al.}(2021)Leng, Wang, Li, Huang, Wang, Wan, Pei, and Wang]{10.1021/acs.inorgchem.1c02621}
Leng,~D.; Wang,~T.; Li,~Y.; Huang,~Z.; Wang,~H.; Wan,~Y.; Pei,~X.; Wang,~J. Plasmonic Bismuth Nanoparticles: Thiolate Pyrolysis Synthesis, Size-Dependent LSPR Property, and Their Oxidation Behavior. \emph{Inorganic Chemistry} \textbf{2021}, \emph{60}, 17258–17267\relax
\mciteBstWouldAddEndPuncttrue
\mciteSetBstMidEndSepPunct{\mcitedefaultmidpunct}
{\mcitedefaultendpunct}{\mcitedefaultseppunct}\relax
\EndOfBibitem
\bibitem[Martínez-Lara \latin{et~al.}(2022)Martínez-Lara, González-Campuzano, and Mendoza]{10.1016/j.photonics.2022.101058}
Martínez-Lara,~D.; González-Campuzano,~R.; Mendoza,~D. Bismuth plasmonics in the visible spectrum using texturized films. \emph{Photonics and Nanostructures - Fundamentals and Applications} \textbf{2022}, \emph{52}, 101058\relax
\mciteBstWouldAddEndPuncttrue
\mciteSetBstMidEndSepPunct{\mcitedefaultmidpunct}
{\mcitedefaultendpunct}{\mcitedefaultseppunct}\relax
\EndOfBibitem
\bibitem[Toudert \latin{et~al.}(2012)Toudert, Serna, and Jiménez~de Castro]{10.1021/jp3065882}
Toudert,~J.; Serna,~R.; Jiménez~de Castro,~M. Exploring the Optical Potential of Nano-Bismuth: Tunable Surface Plasmon Resonances in the Near Ultraviolet-to-Near Infrared Range. \emph{The Journal of Physical Chemistry C} \textbf{2012}, \emph{116}, 20530–20539\relax
\mciteBstWouldAddEndPuncttrue
\mciteSetBstMidEndSepPunct{\mcitedefaultmidpunct}
{\mcitedefaultendpunct}{\mcitedefaultseppunct}\relax
\EndOfBibitem
\bibitem[Ozbay \latin{et~al.}(2020)Ozbay, Ghobadi, Butun, and Turhan-Sayan]{10.1364/ol.45.000686}
Ozbay,~I.; Ghobadi,~A.; Butun,~B.; Turhan-Sayan,~G. Bismuth plasmonics for extraordinary light absorption in deep sub-wavelength geometries. \emph{Optics Letters} \textbf{2020}, \emph{45}, 686\relax
\mciteBstWouldAddEndPuncttrue
\mciteSetBstMidEndSepPunct{\mcitedefaultmidpunct}
{\mcitedefaultendpunct}{\mcitedefaultseppunct}\relax
\EndOfBibitem
\bibitem[Chacon‐Sanchez \latin{et~al.}(2023)Chacon‐Sanchez, de~Galarreta, Nieto‐Pinero, Garcia‐Pardo, Garcia‐Tabares, Ramos, Castillo, Lopez‐Garcia, Siegel, Toudert, Wright, and Serna]{10.1002/adom.202302130}
Chacon‐Sanchez,~F.; de~Galarreta,~C.~R.; Nieto‐Pinero,~E.; Garcia‐Pardo,~M.; Garcia‐Tabares,~E.; Ramos,~N.; Castillo,~M.; Lopez‐Garcia,~M.; Siegel,~J.; Toudert,~J.; Wright,~C.~D.; Serna,~R. Building Conventional Metasurfaces with Unconventional Interband Plasmonics: A Versatile Route for Sustainable Structural Color Generation Based on Bismuth. \emph{Advanced Optical Materials} \textbf{2023}, \emph{12}\relax
\mciteBstWouldAddEndPuncttrue
\mciteSetBstMidEndSepPunct{\mcitedefaultmidpunct}
{\mcitedefaultendpunct}{\mcitedefaultseppunct}\relax
\EndOfBibitem
\bibitem[Tomaszewska \latin{et~al.}(2013)Tomaszewska, Soliwoda, Kadziola, Tkacz-Szczesna, Celichowski, Cichomski, Szmaja, and Grobelny]{10.1155/2013/313081}
Tomaszewska,~E.; Soliwoda,~K.; Kadziola,~K.; Tkacz-Szczesna,~B.; Celichowski,~G.; Cichomski,~M.; Szmaja,~W.; Grobelny,~J. Detection Limits of DLS and UV‐Vis Spectroscopy in Characterization of Polydisperse Nanoparticles Colloids. \emph{Journal of Nanomaterials} \textbf{2013}, \emph{2013}\relax
\mciteBstWouldAddEndPuncttrue
\mciteSetBstMidEndSepPunct{\mcitedefaultmidpunct}
{\mcitedefaultendpunct}{\mcitedefaultseppunct}\relax
\EndOfBibitem
\bibitem[Hendel \latin{et~al.}(2014)Hendel, Wuithschick, Kettemann, Birnbaum, Rademann, and Polte]{10.1021/ac502053s}
Hendel,~T.; Wuithschick,~M.; Kettemann,~F.; Birnbaum,~A.; Rademann,~K.; Polte,~J. In Situ Determination of Colloidal Gold Concentrations with UV–Vis Spectroscopy: Limitations and Perspectives. \emph{Analytical Chemistry} \textbf{2014}, \emph{86}, 11115–11124\relax
\mciteBstWouldAddEndPuncttrue
\mciteSetBstMidEndSepPunct{\mcitedefaultmidpunct}
{\mcitedefaultendpunct}{\mcitedefaultseppunct}\relax
\EndOfBibitem
\bibitem[García~de Abajo(2010)]{10.1103/revmodphys.82.209}
García~de Abajo,~F.~J. Optical excitations in electron microscopy. \emph{Reviews of Modern Physics} \textbf{2010}, \emph{82}, 209–275\relax
\mciteBstWouldAddEndPuncttrue
\mciteSetBstMidEndSepPunct{\mcitedefaultmidpunct}
{\mcitedefaultendpunct}{\mcitedefaultseppunct}\relax
\EndOfBibitem
\bibitem[Horák \latin{et~al.}(2018)Horák, Bukvišová, Švarc, Jaskowiec, Křápek, and Šikola]{10.1038/s41598-018-28037-1}
Horák,~M.; Bukvišová,~K.; Švarc,~V.; Jaskowiec,~J.; Křápek,~V.; Šikola,~T. Comparative study of plasmonic antennas fabricated by electron beam and focused ion beam lithography. \emph{Scientific Reports} \textbf{2018}, \emph{8}, 9640\relax
\mciteBstWouldAddEndPuncttrue
\mciteSetBstMidEndSepPunct{\mcitedefaultmidpunct}
{\mcitedefaultendpunct}{\mcitedefaultseppunct}\relax
\EndOfBibitem
\bibitem[Horák and Šikola(2020)Horák, and Šikola]{10.1016/j.ultramic.2020.113044}
Horák,~M.; Šikola,~T. Influence of experimental conditions on localized surface plasmon resonances measurement by electron energy loss spectroscopy. \emph{Ultramicroscopy} \textbf{2020}, \emph{216}, 113044\relax
\mciteBstWouldAddEndPuncttrue
\mciteSetBstMidEndSepPunct{\mcitedefaultmidpunct}
{\mcitedefaultendpunct}{\mcitedefaultseppunct}\relax
\EndOfBibitem
\bibitem[Křápek \latin{et~al.}(2019)Křápek, Konečná, Horák, Ligmajer, St\"{o}ger-Pollach, Hrtoň, Babocký, and Šikola]{10.1515/nanoph-2019-0326}
Křápek,~V.; Konečná,~A.; Horák,~M.; Ligmajer,~F.; St\"{o}ger-Pollach,~M.; Hrtoň,~M.; Babocký,~J.; Šikola,~T. Independent engineering of individual plasmon modes in plasmonic dimers with conductive and capacitive coupling. \emph{Nanophotonics} \textbf{2019}, \emph{9}, 623–632\relax
\mciteBstWouldAddEndPuncttrue
\mciteSetBstMidEndSepPunct{\mcitedefaultmidpunct}
{\mcitedefaultendpunct}{\mcitedefaultseppunct}\relax
\EndOfBibitem
\bibitem[Sano \latin{et~al.}(2005)Sano, Satoh, Chiba, Shinohara, Okamoto, Serizawa, Nakashima, and Omae]{10.1539/joh.47.242}
Sano,~Y.; Satoh,~H.; Chiba,~M.; Shinohara,~A.; Okamoto,~M.; Serizawa,~K.; Nakashima,~H.; Omae,~K. A 13‐Week Toxicity Study of Bismuth in Rats by Intratracheal Intermittent Administration. \emph{Journal of Occupational Health} \textbf{2005}, \emph{47}, 242–248\relax
\mciteBstWouldAddEndPuncttrue
\mciteSetBstMidEndSepPunct{\mcitedefaultmidpunct}
{\mcitedefaultendpunct}{\mcitedefaultseppunct}\relax
\EndOfBibitem
\bibitem[Griffith \latin{et~al.}(2021)Griffith, Li, Werrett, Andrews, and Sun]{10.1039/D0CS00031K}
Griffith,~D.~M.; Li,~H.; Werrett,~M.~V.; Andrews,~P.~C.; Sun,~H. Medicinal chemistry and biomedical applications of bismuth-based compounds and nanoparticles. \emph{Chemical Society Reviews} \textbf{2021}, \emph{50}, 12037–12069\relax
\mciteBstWouldAddEndPuncttrue
\mciteSetBstMidEndSepPunct{\mcitedefaultmidpunct}
{\mcitedefaultendpunct}{\mcitedefaultseppunct}\relax
\EndOfBibitem
\bibitem[Waxenegger \latin{et~al.}(2015)Waxenegger, Trügler, and Hohenester]{10.1016/j.cpc.2015.03.023}
Waxenegger,~J.; Trügler,~A.; Hohenester,~U. Plasmonics simulations with the MNPBEM toolbox: Consideration of substrates and layer structures. \emph{Computer Physics Communications} \textbf{2015}, \emph{193}, 138--150\relax
\mciteBstWouldAddEndPuncttrue
\mciteSetBstMidEndSepPunct{\mcitedefaultmidpunct}
{\mcitedefaultendpunct}{\mcitedefaultseppunct}\relax
\EndOfBibitem
\bibitem[Werner \latin{et~al.}(2009)Werner, Glantschnig, and Ambrosch-Draxl]{10.1063/1.3243762}
Werner,~W. S.~M.; Glantschnig,~K.; Ambrosch-Draxl,~C. Optical Constants and Inelastic Electron-Scattering Data for 17 Elemental Metals. \emph{Journal of Physical and Chemical Reference Data} \textbf{2009}, \emph{38}, 1013–1092\relax
\mciteBstWouldAddEndPuncttrue
\mciteSetBstMidEndSepPunct{\mcitedefaultmidpunct}
{\mcitedefaultendpunct}{\mcitedefaultseppunct}\relax
\EndOfBibitem
\end{mcitethebibliography}

\newpage
\section{Supplementary Information}

\setcounter{figure}{0}
\setcounter{table}{0}
\renewcommand{\thefigure}{S\arabic{figure}} 
\renewcommand{\thetable}{S\arabic{table}}

\begin{figure}
   \centering
   \includegraphics[width=0.9\linewidth]{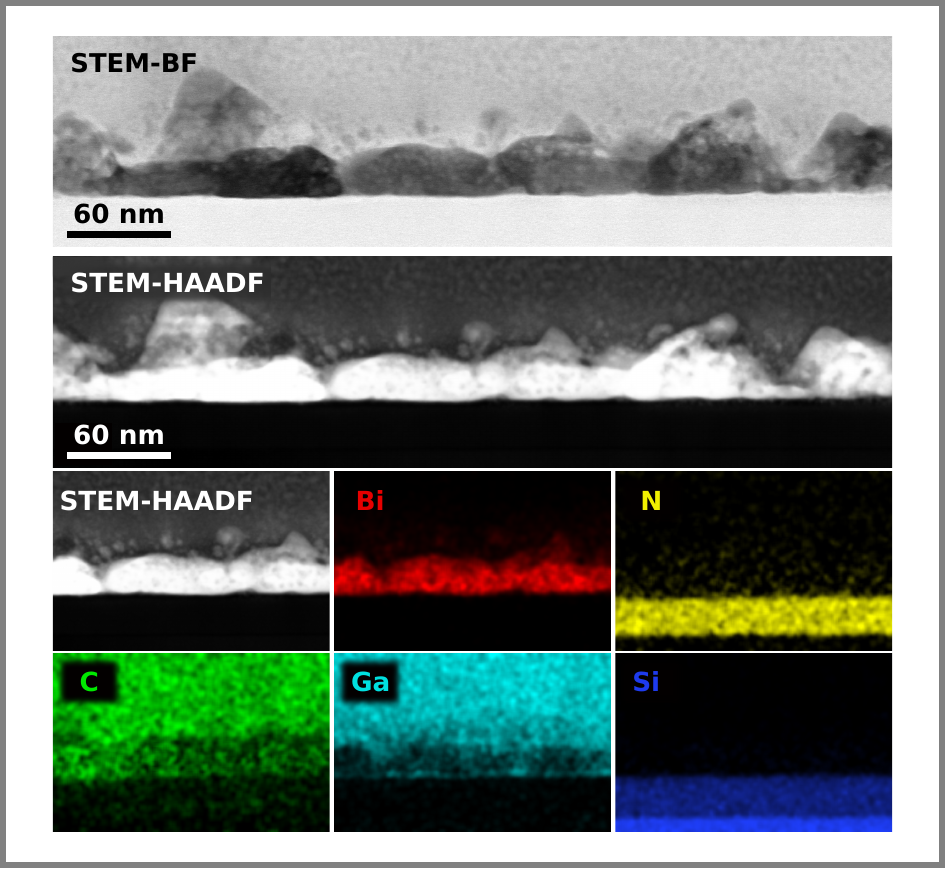}
   \caption{Material analysis of the \SI{30}{\nano\meter} thick bismuth thin film deposited on a \SI{30}{\nano\meter} thick silicon nitride membrane performed by STEM EDX (energy-dispersive X-ray spectroscopy) mapping. The cross-sectional view show the pronounced roughness of the bismuth layer. The composition is from top to bottom the following: lamella protection layer (carbon and gallium), bismuth layer, silicon nitride membrane, and silicon supportive frame.}
\end{figure}

\begin{figure}
   \centering
   \includegraphics[width=1\linewidth]{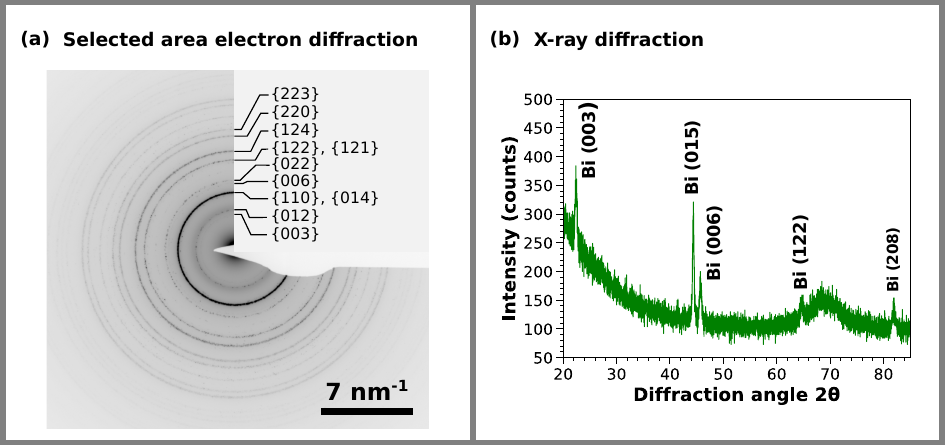}
   \caption{Diffraction analysis of the \SI{30}{\nano\meter} thick bismuth thin film deposited on a \SI{30}{\nano\meter} thick silicon nitride membrane performed (a) by selective area electron diffraction (SAED) in the TEM and (b) by X-ray diffraction (XRD). Both diffractograms contain no signs of the presence of bismuth oxide, therefore the bismuth layer is not oxidized.}
\end{figure}

\begin{figure}
   \centering
   \includegraphics[width=1\linewidth]{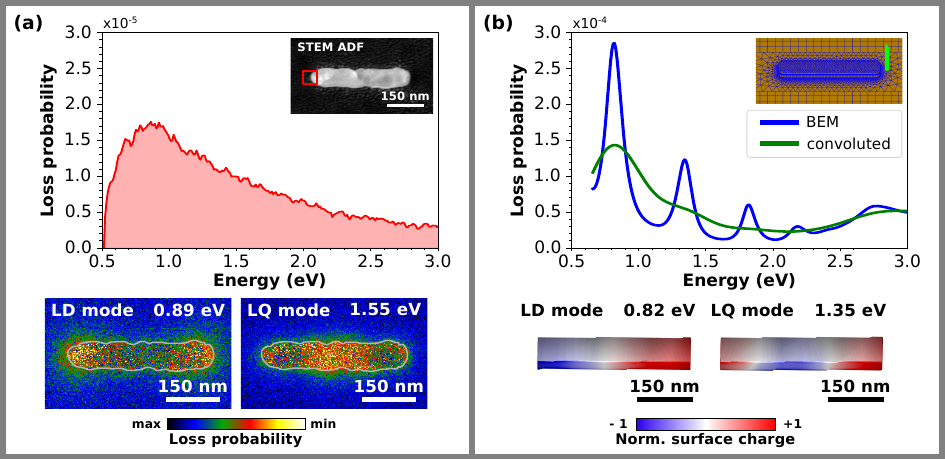}
   \caption{Plasmonic response of a \SI{410}{\nano\meter} long bar-shaped bismuth antenna. (a) STEM ADF micrograph of the bar, EEL spectrum, and energy filtered loss probability maps at the energy of \SI{0.89}{\electronvolt} and \SI{1.55}{\electronvolt} corresponding to the longitudinal dipole (LD) and longitudinal quadrupole (LQ) mode. (b) Numerical simulation of the EEL specrtum that is further convolved with a Gaussian function to reproduce the instrumental broadening of the peaks. The LD and LQ mode is visualized by the normalized surface charge distributions at the peak energy of these modes (\SI{0.82}{\electronvolt} and \SI{1.35}{\electronvolt}). The theory represented by numerical simulations matches well the experiment.}
\end{figure}

\begin{figure}
   \centering
   \includegraphics[width=1\linewidth]{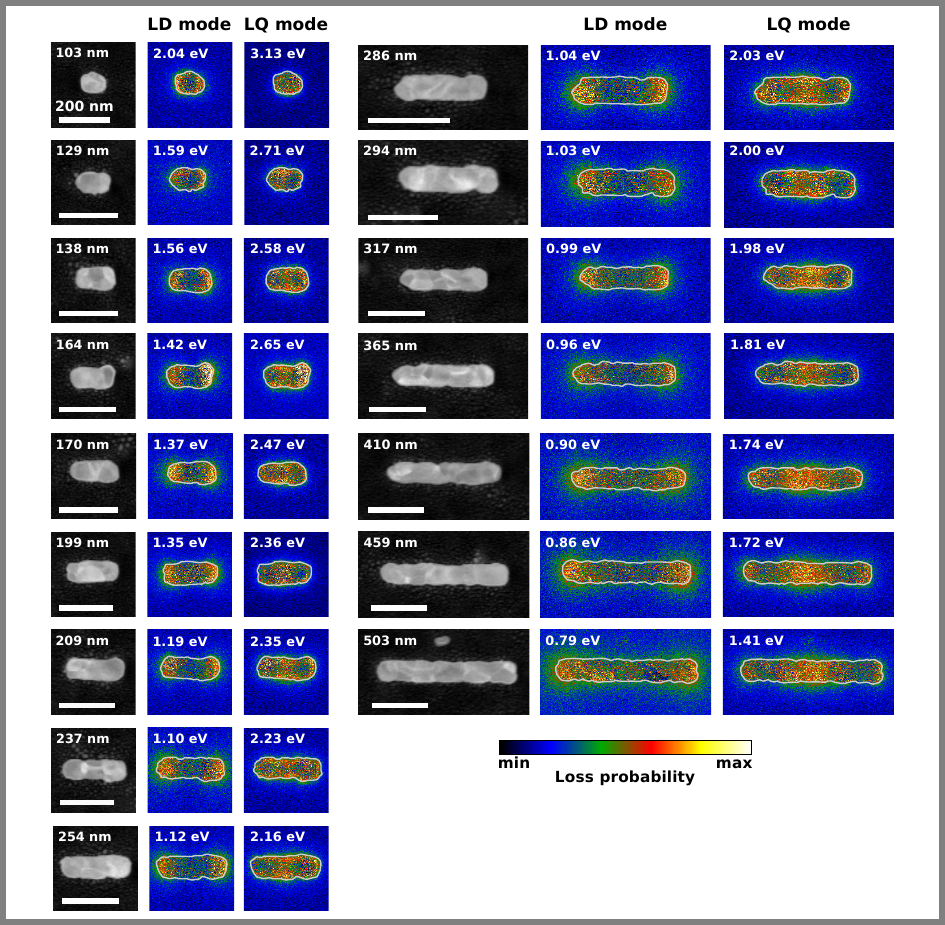}
   \caption{Energy filtered loss probability maps of the bar-shaped bismuth antennas at the energies corresponding to their longitudinal dipole (LD) and longitudinal quadrupole (LQ) modes. The scalebars are \SI{200}{\nano\meter} long.}
\end{figure}

\begin{figure}
   \centering
   \includegraphics[width=1\linewidth]{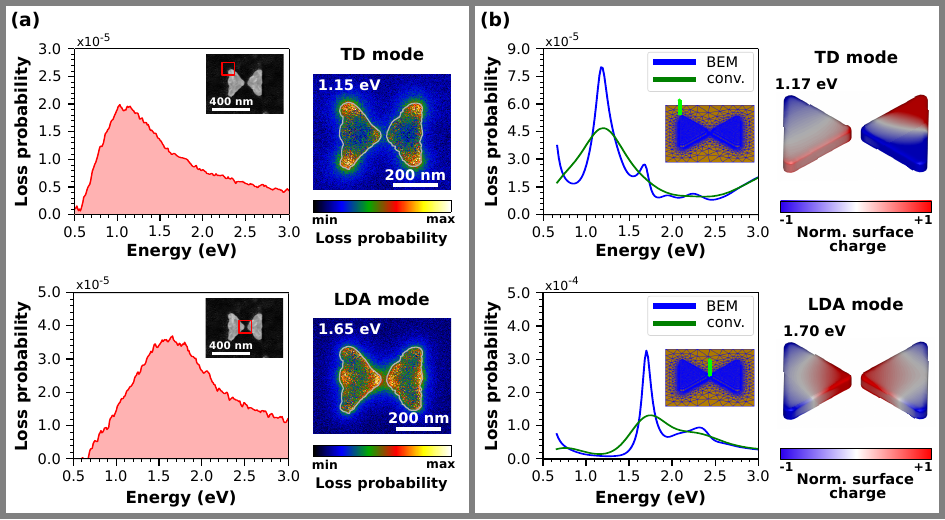}
   \caption{Plasmonic response of a \SI{288}{\nano\meter} wide bowtie bismuth antenna. (a) STEM ADF micrograph of the bowtie, EEL spectra recorded at the outer corner and in the gap of the bowtie, and energy filtered loss probability maps at the energy of \SI{1.15}{\electronvolt} and \SI{1.65}{\electronvolt} corresponding to the transverse dipole (TD) and longitudinal dipole antibonding (LDA) mode. (b) Numerical simulations of the EEL spectra at the outer corner and in the gap of the bowtie that are further convolved with a Gaussian function to reproduce the instrumental broadening of the peaks. The TD and LDA mode is visualized by the normalized surface charge distributions at the peak energy of these modes (\SI{1.17}{\electronvolt} and \SI{1.70}{\electronvolt}). The theory represented by numerical simulations matches well the experiment.}
\end{figure}

\begin{figure}
   \centering
   \includegraphics[width=1\linewidth]{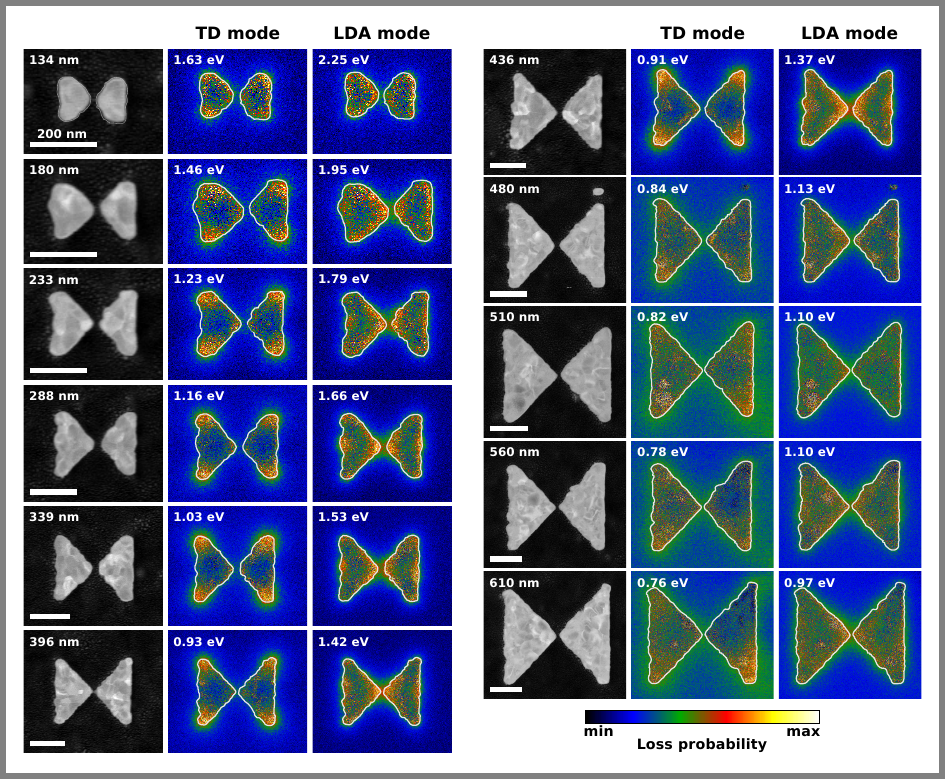}
   \caption{Energy filtered loss probability maps of the bowtie bismuth antennas at the energies corresponding to their transverse dipole (TD) and longitudinal dipole antibonding (LDA) modes. The scalebars are \SI{200}{\nano\meter} long.}
\end{figure}

\begin{figure}
   \centering
   \includegraphics[width=1\linewidth]{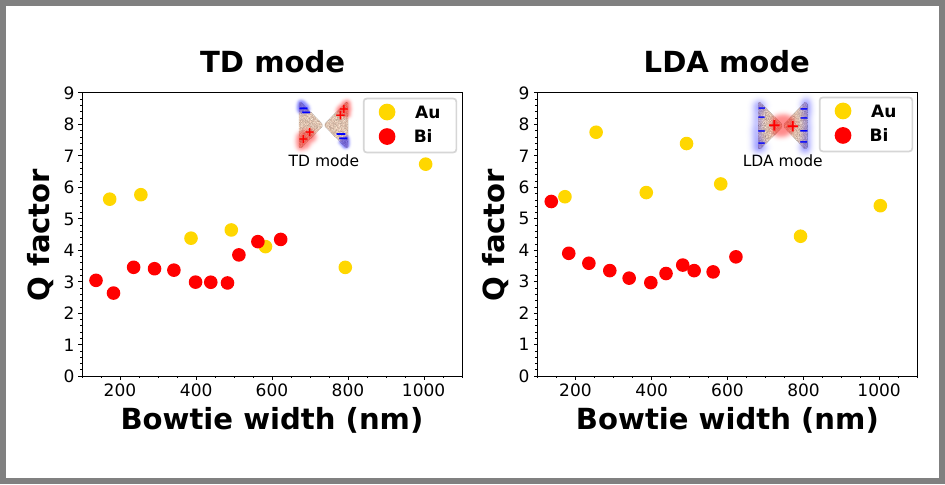}
   \caption{Comparison of Q factors for the TD (a) and LDA (b) modes in the bowtie bismuth and gold antennas.}
\end{figure}

\end{document}